\documentclass[%
 aip,
  pof,     
 amsmath,amssymb,
 reprint,%
 groupedaddress,
]{revtex4-1}

\usepackage{graphicx}
\usepackage{dcolumn}
\usepackage{bm}
\usepackage{verbatim}
\usepackage{color}

\begin{document}

\title{Bubble-Driven Inertial Micropump}

\author{Erik D. Torniainen}
 \email{erik.torniainen@hp.com}
 \affiliation{Hewlett-Packard Company, Imaging and Printing Division, Corvallis, Oregon 97330 USA} 

\author{Alexander N. Govyadinov}
 \affiliation{Hewlett-Packard Company, Imaging and Printing Division, Corvallis, Oregon 97330 USA} 
 
\author{David P. Markel}
 \affiliation{Hewlett-Packard Company, Imaging and Printing Division, Corvallis, Oregon 97330 USA}  
 
\author{Pavel E. Kornilovitch}
 \affiliation{Hewlett-Packard Company, Imaging and Printing Division, Corvallis, Oregon 97330 USA} 

\date{\today}  

\begin{abstract}

The fundamental action of the bubble-driven inertial micropump is investigated. The pump has no moving parts and consists of a thermal resistor placed asymmetrically within a straight channel connecting two reservoirs. Using numerical simulations, the net flow is studied as a function of channel geometry, resistor location, vapor bubble strength, fluid viscosity, and surface tension. Two major regimes of behavior are identified: axial and non-axial. In the axial regime, the drive bubble either remains inside the channel or continues to grow axially when it reaches the reservoir. In the non-axial regime the bubble grows out of the channel and in all three dimensions while inside the reservoir. The net flow in the axial regime is parabolic with respect to the hydraulic diameter of the channel cross-section but in the non-axial regime it is not. From numerical modeling, it is determined that the net flow is maximal when the axial regime crosses over to the non-axial regime. To elucidate the basic physical principles of the pump, a phenomenological one-dimensional model is developed and solved. A linear array of micropumps has been built using silicon-SU8 fabrication technology, and semi-continuous pumping across a 2 mm-wide channel has been demonstrated experimentally. Measured variation of the net flow with fluid viscosity is in excellent agreement with simulation results.           

\end{abstract}


\maketitle

\section{\label{sec:one}
Introduction
}

Adoption of microfluidic devices for biomedical, chemical, engineering, and other applications~\cite{Yole2011} depends on the successful development of microfluidic components: mixers, filters, heaters, valves, pumps, and sensors. Such components must be simple and reliable, and ideally produced by scalable fabrication methods to keep the cost of the technology and production low. This is not easily achievable for complex active components, specifically the micropump. 

A large number of different pumping principles have been described in the literature,~\cite{Woias2001,Laser2004,Oh2006,Nisar2008,Amirouche2009,Au2011} but many of the current approaches have downsides. Pumps with active valves contain moving parts so they may not be very reliable or easy to fabricate. Passive capillary pumps require no external power but may not be adaptable to a wide range of fluids. Electro-osmotic pumps do not require any moving parts but work only with electrically conductive fluids and need high voltages. Other micropump technologies exist and are being developed, but ultimately there are few which are reliable, small, adaptable to a wide range of fluids, and mass producible. According to Oh and Ahn~\cite{Oh2006} ``the successful miniaturization and commercialization of fully integrated microfluidic systems have been delayed due to the lack of reliable microfluidic components, i.e., micropumps and microvalves. Therefore, even though much attention has been paid to the development of the microfluidic components, they are still the most difficult task.'' Stone~\cite{Stone2009} expressed a similar opinion.        

One type of mechanical actuation available at microscale is the explosive power of a superheated fluid vapor bubble. In such a system, a small volume of liquid is heated well above the boiling point by a local heater while the rest of the system remains essentially at room temperature. The pressure inside the bubble can reach many atmospheres which then efficiently pushes the fluid through a network of adjacent microchannels. This driving principle is behind the commercially successful Thermal Inkjet (TIJ) printing technology.~\cite{Stasiak2012} In TIJ, the vapor bubble ejects a droplet of ink into the ambient environment via an open nozzle. The important question is whether the same driving principle can be adapted to move fluids around in {\em closed} systems. One approach is to scale down a peristaltic pump while using several thermal bubbles in place of the traditional mechanical actuators,~\cite{Jun1996} however the pump was too inefficient for practical applications.       

A different approach was described by Yuan and Prosperetti.~\cite{Yuan1999b} In this method, fluid is driven by a {\em single} bubble, repeatedly expanding and collapsing inside a narrow microchannel that connects two large reservoirs. If the bubble is created in the middle of the channel the net effect is zero by symmetry. However, if the bubble is created closer to one end of the channel and expands rapidly enough, non-linearity is generated in the fluid motion which results in a net flow from the short arm to the long arm of the channel. Yuan and Prosperetti attributed the effect to differences in the mechanical {\em inertia} of the two columns of fluids. Later, Yin and Prosperetti demonstrated the pumping action experimentally by using the electrical conductivity of the fluid~\cite{Yin2005a} and resistive heaters~\cite{Yin2005b} to generate the bubbles. Pumping in microtubes using laser-induced cavitation to create bubbles has also been examined.\cite{Wang2004,Dijkink2008}

The motivation behind this work is to conduct a deep study of the {\em bubble-driven inertial pump} and develop an understanding of the operating space including the effects of geometry and fluid properties. Resistively driven inertial pumps, supporting electronics, and microfluidic networks can be made by utilizing the same fabrication processes used to manufacture TIJ printheads. Understanding the pump operation for realistic sizes, geometries, energies, and fluid parameters is critical when leveraging these scalable technologies to design future devices that contain pump densities as high as thousands per square centimeter.

This paper begins with a discussion of the physics behind the pumping mechanism (Section~\ref{sec:two}). A simple dynamical model is introduced and solved, and then the main properties of the pump are illustrated using the model. In Section~\ref{sec:three} the results of comprehensive Computational Fluid Dynamics (CFD) modeling are presented. The net flow is calculated for different channel cross-sections, resistor locations, and fluid viscosities and surface tensions. Two different bubble expansion regimes, axial and non-axial, are identified and discussed. The self-similar nature of the flow in the axial regime is established. Finally, conditions for maximizing the net flow are given and pump curves calculated. Section~\ref{sec:four} describes experimental demonstration of inertial pumping and its correlation to CFD analysis.

\section{\label{sec:two}
Operating principle of the inertial pump
}

\begin{figure}[t]
\includegraphics[width=0.48\textwidth]{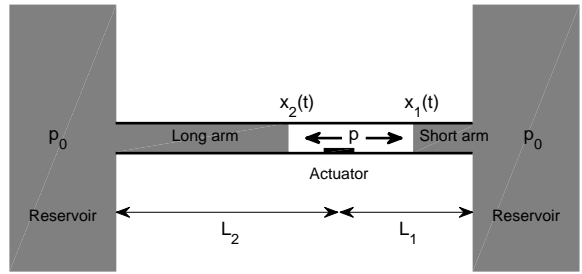}
\vspace{-0.5cm}
\caption{Schematics of inertial micropump. The actuator is at the origin of the coordinate axis, $x = 0$.}   
\label{fig:one}
\end{figure}

To elucidate the physical principles behind inertial pumping it is useful to consider a simplified one-di\-men\-si\-o\-nal model illustrated in Fig.~\ref{fig:one}. The pump contains an actuator (thermal resistor) located in a microchannel that separates two reservoirs at ambient pressure $p_0$. The side of the channel where the actuator is closer to the reservoir is called the short arm and the other side which is further away from the reservoir is called the long arm. The actuator creates a high pressure region (a model vapor bubble) of pressure $p > p_0$, which induces outward flow in both arms. This phase will be referred to as ``bubble expansion.'' As the bubble expands, the pressure drops below $p_0$ and outward flow is replaced by inward flow or ``bubble collapse.'' The two inward flows collide at a point that is in general not the same as the starting point of the bubble expansion. This is equivalent to a net flow as a result of the expansion-collapse cycle. In addition, the mechanical momenta of the two colliding fluid columns may be different. In this case, the fluid will continue to flow even after the collapse, until the total momentum is dissipated via external and internal friction. This phase will be referred to as ``post-collapse.'' The total net flow of a pump cycle is the sum of the expansion-collapse and post-collapse contributions.         

\subsection{\label{sec:twoone}
Momentum balance
}

Further insight into the pump operation can be gain\-ed by analyzing momentum balance. The arguments given below differ somewhat from those of Yuan and Prosperetti,\cite{Yuan1999b} but they lead to the same model as given by Yin and Prosperetti.~\cite{Yin2005a,Yin2005b} In the one-dimensional model of Fig.~\ref{fig:one}, the bubble size is defined by the end coordinates $x_1(t)$ and $x_2(t)$. The same variables completely determine the dynamics of the two fluid columns. In particular, the velocities are given by $v_1(t) = \dot{x}_1(t)$ and $v_2(t) = \dot{x}_2(t)$ for the right and left arms, respectively. The mechanical momenta are given by
\begin{eqnarray}
Q_1(t) & = &  M_1(t) v_1(t) = \rho A ( L_1 - x_1 ) \, \dot{x}_1 \: ,
\label{eq:twentyone} \\
Q_2(t) & = &  M_2(t) v_2(t) = \rho A ( L_2 + x_2 ) \, \dot{x}_2 \: ,
\label{eq:twentytwo} 
\end{eqnarray}
where $\rho$ is the fluid mass density and $A$ is the cross-sectional area of the channel. 

It is essential that the fluid momentum of the reservoir is much smaller than that of the channel as long as the reservoir cross-section is much larger than $A$. Under these conditions, the reservoir momentum can be neglected. Within the simplified model, the momentum of any fluid exiting from the channel into a reservoir is regarded as a {\em net loss} of momentum in the balance equation. Similarly, the momentum of any fluid element entering the channel from a reservoir is a {\em net gain} of momentum. (In the full three-dimensional modeling of Section~\ref{sec:three} the reservoir momentum is properly taken into account. There the reservoir momentum does not get ``lost'' but instead creates vortices near the channel end.) Consider, for example, outward flow of the right fluid column (the short arm). Within a time increment $dt$ the momentum is changed because of two reasons: (i) External forces produce a change $F_1 \, dt$, which is positive or negative depending on the sign of $F_1$; (ii) A mass element $dM_1 = \rho A v_1 \, dt$ exits into the right reservoir. As a result, momentum $\rho A \dot{x}^2_1 \, dt$ is ``lost'', and has to be included in the balance as a decrement. The balance equation reads
\begin{equation}
Q_1(t + dt) = Q_1(t) - \rho A \dot{x}^2_1 \, dt + F_1 \, dt \: .
\label{eq:twentythree} 
\end{equation}
Substituting here Eq.~(\ref{eq:twentyone}) and rearranging terms, one obtains
\begin{equation}
\rho A ( L_1 - x_1 ) \, \ddot{x}_1 = F_1 \: .
\label{eq:twentyfour} 
\end{equation}
During the collapse phase, fluid enters the channel from the right reservoir and contributes a momentum $\rho A \dot{x}^2_1 \, dt$. However, since the right column is now moving to the left, this increment has to be taken with a {\em negative} sign. This again results in Eq.~(\ref{eq:twentythree}); however, $F_1$ now has a different sign and magnitude. One concludes that Eq.~(\ref{eq:twentyfour}) describes both the expansion and collapse phases of the right column. Equation~(\ref{eq:twentyfour}) is equivalent to the one given by Yin and Prosperetti.~\cite{Yin2005a,Yin2005b}

Consider now the dynamics of the left column (the long arm) in Fig.~\ref{fig:one}. During expansion, {\em negative} momentum $\rho A \dot{x}^2_2 \, dt$ is ``lost'' to the left reservoir. Therefore it represents a net {\em gain} of momentum and must be included in the balance with a positive sign:  
\begin{equation}
Q_2(t + dt) = Q_2(t) + \rho A \dot{x}^2_2 \, dt + F_2 \, dt \: .
\label{eq:twentyfive} 
\end{equation}
Together with the definition (\ref{eq:twentytwo}), this leads to the dynamic equation
\begin{equation}
\rho A ( L_2 + x_2 ) \, \ddot{x}_2 = F_2 \: .
\label{eq:twentysix} 
\end{equation}
During collapse, fluid enters from the left reservoir with a positive velocity and {\em adds} to the balance a momentum $\rho A \dot{x}^2_2 \, dt$. This term has to be taken with a plus sign, which leads again to equation (\ref{eq:twentysix}). 

Thus in the simplified model, the fluid is described by a Newton-like equation with a variable mass. However, the equation does not explicitly contain the time derivative of the mass, since the latter is precisely cancelled by the momentum lost to the reservoir.

\subsection{\label{sec:twotwo}
Forces and dimensionless parameters
}

To complete the model, external forces $F$ need to be specified. In this work, three types of forces are included: (i) The driving force that is proportional to the pressure difference between the bubble and the reservoirs $(p - p_0)$ and the channel area $A$, (ii) The viscous force that is chosen to be proportional to the length of the fluidic column and to its velocity with a coefficient $\kappa$. (iii) Surface tension with coefficient $\sigma$. The dynamic equations assume the form         
\begin{eqnarray}
\rho A ( L_1 - x_1 ) \, \ddot{x}_1 \! + \! \kappa \, ( L_1 - x_1 ) \, \dot{x}_1 & = & 
( p - p_0 ) A \! - \! \frac{4 \sigma A}{D_h} ,
\label{eq:one} \\
\rho A ( L_2 + x_2 ) \, \ddot{x}_2 \! + \! \kappa \, ( L_2 + x_2 ) \, \dot{x}_2 & = & 
( p_0 - p ) A \! + \! \frac{4 \sigma A}{D_h} .
\label{eq:two} 
\end{eqnarray}
Note that $\kappa$ has the dimensionality of dynamic viscosity but its numerical value is typically larger than the bulk viscosity of the participating fluid because $\kappa$ approximates the entire viscous effect. The bubble pressure $p$ is larger than the external pressure $p_0$ during the expansion and smaller than $p_0$ during the collapse. The surface tension term is defined via the hydraulic diameter $D_h$  
\begin{equation}
D_h = \frac{4 A}{P} \: ,
\label{eq:eightone} 
\end{equation}
where $P$ is the perimeter of the channel cross-section. $D_h$ represents the equivalent circular diameter of the rectangular cross-section.~\cite{White1991}

It is instructive to transform the dynamic equations into dimensionless form. One way to achieve that is to introduce characteristic time $t_0$ and distance $a_0$. Then $a_0/t_0$ is an estimate of the typical velocity. The latter may be used to construct useful dimensionless numbers. For the geometries studied in the present paper, single pumping events take microseconds and occur on the scale of microns. As a first-order approximation, it is convenient to take $t_0 = 1$ $\mu$s and $a_0 = 1$ $\mu$m. Using $\bar{x}_1 \equiv x_1/a_0$, $\bar{t} \equiv t/t_0$ and dividing by $\rho A a^2_0/t^2_0$ one obtains from Eq.~(\ref{eq:one}) 
\begin{equation}
\left( \frac{L_1}{a_0} - \bar{x}_1 \right) \bar{x}''_1 + 
\alpha \left( \frac{L_1}{a_0} - \bar{x}_1 \right) \bar{x}'_1 = 
\beta \left( \frac{p}{p_0} - 1 \right) -  4 \gamma \: ,
\label{eq:seventeen}
\end{equation}
and a similar equation from Eq.~(\ref{eq:two}). The emerged dimensionless combinations are
\begin{eqnarray}
\alpha & = & \frac{\kappa t_0}{\rho A}           \sim 0.1   \: ,
\label{eq:thirtyone}   \\
\beta  & = & \frac{p_0 t^2_0}{\rho a^2_0}        \sim 100   \: ,
\label{eq:thirtytwo}   \\
\gamma & = & \frac{\rho D_h a^2_0}{\sigma t^2_0} \sim 0.1   \: .
\label{eq:thirtythree}
\end{eqnarray}
The numerical estimates have been obtained with $\rho = 1$ g/cm$^3$, $\sigma = 100$ dyn/cm, $\kappa = 10$ cP, $p_0 = 1$ atm, $D_h = 10$ $\mu$m, and $A = D^2_h$. One can see that the pressure force is much larger than the viscous and surface tension forces. Therefore during the expansion and collapse phases the viscous force and surface tension force are negligible, and the pressure force is balanced primarily by inertia.   

One should note that the parameter $\alpha$ is roughly an inverse of the Reynolds number. By choosing $(A/a_0)$ being the characteristic scale, and $U = a_0/t_0$ a characteristic fluid velocity, it can be rewritten as $\alpha = \kappa/[\rho U (A/a_0)] = 1/{\rm Re}$. Likewise, $\gamma$ is an inverse of the Weber number: $\gamma = \rho D_h U^2/\sigma = 1/{\rm We}$. Proper definitions of Re and We for the problem at hand require a more rigorous derivation of the characteristic velocity from the details of fluid dynamics. This will be done in Section~\ref{sec:three}.        

The force balance changes in the post-collapse phase. This phase begins after the two columns collide at a point $x_c$ at a time $t_c$. The fluid fills the channel completely and moves as a whole with a constant mass $\rho A (L_1 + L_2)$. At this moment, the surface tension and pressure forces disappear. The fluid inertia is balanced by the viscous force. The latter dissipates the mechanical momentum to a complete stop. The motion during this phase can still be described by the time evolution of the collision point $x(t)$. The dynamic equation reads:    
\begin{equation}
\rho A ( L_1 + L_2 ) \, \ddot{x} + \kappa \, ( L_1 + L_2 ) \, \dot{x} = 0 \: ,
\label{eq:three} 
\end{equation}
with the solution  
\begin{equation}
x(t) - x_c = \frac{\rho A}{\kappa} \, v(t_c) \left[ 1 - e^{- \frac{\kappa}{\rho A} ( t - t_c ) } \right]  \: .
\label{eq:four} 
\end{equation}
Here $v(t_c)$ is the velocity at the beginning of the post-collapse phase. $v(t_c)$ can be found from the total momentum at the end of the collapse phase. The value $x(\infty)$ is the measure of the total net flow in one pump stroke. The net {\em volume} moved in one stroke is given by 
\begin{equation}
V_1 = A x(\infty) = \frac{\rho A^2 v(t_c)}{\kappa} + A x_c \: .
\label{eq:five} 
\end{equation}
\begin{figure}[t]
\includegraphics[width=0.48\textwidth]{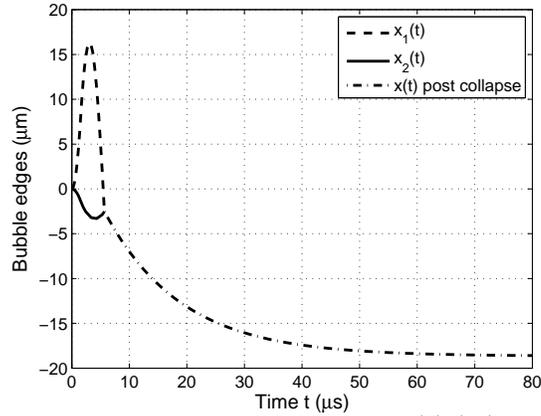}
\vspace{-0.5cm}
\caption{Numerical solution of the one-dimensional dynamical model (\ref{eq:one})-(\ref{eq:three}). The parameters are: $L_1$ = 30 $\mu$m, $L_2$ = 170 $\mu$m, $A$ = 20 $\times$ 20 $\mu{\rm m}^2$, $\rho$ = 1.0 g/cm$^3$, effective viscosity $\kappa$ = 30 cP, $\sigma = 50$ dyn/cm, $p$ = 3 atm, $p_0$ = 1 atm, time duration of high pressure = 1.5 $\mu$s.}   
\label{fig:two}
\end{figure}

\subsection{\label{sec:twothree}
Numerical solution
}

The equations (\ref{eq:one})-(\ref{eq:two}) have been solved numerically using MATLAB. A typical result, corresponding to systems studied in Section~\ref{sec:three}, is shown in Fig.~\ref{fig:two}. The short and long arms are 30 $\mu$m and 170 $\mu$m in length and other parameters are listed in the caption. The high pressure phase lasts 1.5 $\mu$s with time-independent pressure of 3 atmospheres. After the positive pressure is replaced by a negative pressure of 1 atmosphere the bubble continues to expand by inertia until expansion stops and the bubble begins to collapse. Notice that due to differences in inertia, the collapse phase of the short arm starts earlier (at $\approx$ 3.0 $\mu$s) than that of the long arm (at $\approx$ 4.0 $\mu$s). The collapse phase ends at $\approx$ 5.6 $\mu$s when the two columns collide at the point $x_c = -2.5$ $\mu$m to the left from the actuator. (The curves cross below zero). Since the mechanical momentum of the short arm is larger, after the collapse the entire fluid column continues to move to the left in accordance with Eq.~(\ref{eq:three}). The additional displacement of the post-collapse phase is $-16.1$ $\mu$m. The total {\em volume} moved in a single event is 7.4 picoliters.  

If the fluid mass were assumed to be independent of the bubble size during the expansion-collapse phase the net flow would be zero. If the mass term were completely omitted from the dynamic equations (i.e. the flow would be determined by the balance of the pressure and viscous forces only), the net flow would also be zero. The conclusion is the net flow is caused by variation of the mass (inertness) of the fluid columns with their positions. As the two opposite flows reverse direction near the point of maximal expansion, the short column decelerates faster and starts flowing back faster than the long column. As a result, {\em the net flow will be from the short arm toward the long arm of the channel}. Since the effect is caused by differences in inertial properties, the term ``inertial pumping'' is justified. The difference in inertia is zero for symmetric placement of the actuator and increases as the actuator is shifted from the channel center. Therefore one expects the pumping to be stronger for more asymmetric placement of the actuator. The actuator location is defined by parameter $R$ as follows
\begin{equation}
R = \frac{L_2}{L_1 + L_2} \: .
\label{eq:twenty} 
\end{equation}
The symmetric placement of the actuator corresponds to $R = 0.5$ and one expects zero net flow. The net flow should increase as $R$ deviates from 0.5. Throughout this paper, the net flow in picoliters (pL) will be plotted as a function of $R$. It should be kept in mind that for $0.5 \leq R \leq 1.0$ (the resistor is shifted to the right from the symmetric position), the net flow will be to the {\em left}, (i.e. from the short to the long arm). What is plotted is the {\em absolute value} of the net flow from right to left.    

\begin{figure}[t]
\includegraphics[width=0.48\textwidth]{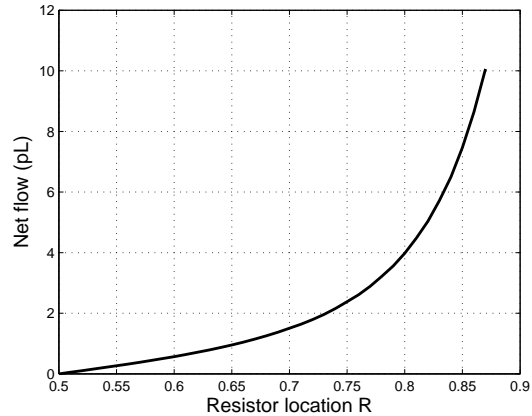}
\vspace{-0.5cm}
\caption{Net flow in model (\ref{eq:one})-(\ref{eq:three}) as a function of resistor location for the parameter set of Fig.~\ref{fig:two}. The total channel length is $L_1 + L_2 = 200$ $\mu$m.}   
\label{fig:twotwo}
\end{figure}

Figure~\ref{fig:twotwo} shows the net flow dependence on $R$ calculated within the one-dimensional model (\ref{eq:one})-(\ref{eq:three}). The net flow grows linearly for small asymmetries and then faster as $R$ approaches 1. The increase continues until the bubble starts breaking into the reservoir on the shorter arm, and the one-dimensional model becomes invalid.   

While providing useful insights, the one-dimensional model does not capture several important aspects of the inertial pump operation such as the time-dependence of the vapor pressure in the bubble,~\cite{Sun2009} transverse velocity gradients in the channel, and three-dimensional flow in the reservoirs. These limitations can be overcome with full three-dimensional fluidic modeling as presented in the next Section.

\section{\label{sec:three}
Computational Fluid Dynamics modeling
}

The Computational Fluid Dynamics (CFD) code used for simulations in this paper is an internal software application developed within Hewlett-Packard to model incompressible fluid flows with free surfaces, particularly focused on the application of ink jet printing. This code is an explicit, transient code which solves the incompressible Navier-Stokes equations utilizing the Volume of Fluid method to reproduce free surfaces.~\cite{Knight1991} Extensive comparisons to experiment have refined this CFD code, particularly with respect to representing ink jet drop ejection.~\cite{Knight1985} The commercial program Tecplot is used to aid with visualization of the solution.

\begin{figure}[t]
\includegraphics[width=0.48\textwidth]{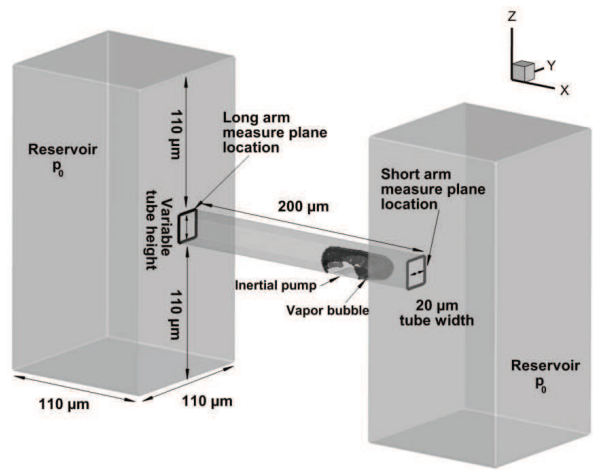}
\caption{Schematic of two-reservoir pump geometry. The channel containing a pump resistor is 200 $\mu$m long and 20 $\mu$m wide with variable height.}
\label{fig:three}
\end{figure}

The computational grid is one of many user defined CFD parameters and for this particular geometry a 1.0 $\mu$m grid, uniform in all coordinate directions, was utilized. Figure~\ref{fig:three} shows the overall geometry and dimensions of the prototypical geometry used in this study. The geometry is analogous to the geometry in Section~\ref{sec:two} of the paper and contains a rectangular channel with a thermal resistor (actuator) located between two reservoirs. Static pressure boundary conditions were applied to the reservoir edges and no-slip boundary conditions were applied to channel walls. Interfaces between liquid and gas had no tangential stress and interfacial normal stress accommodated pressure and surface curvature in a typical fashion. Extensive simulations were performed to examine the appropriate reservoir size, boundary conditions, and grid resolution to minimize their effect on the simulation results.   

\begin{figure*}[t]
\includegraphics[width=0.90\textwidth]{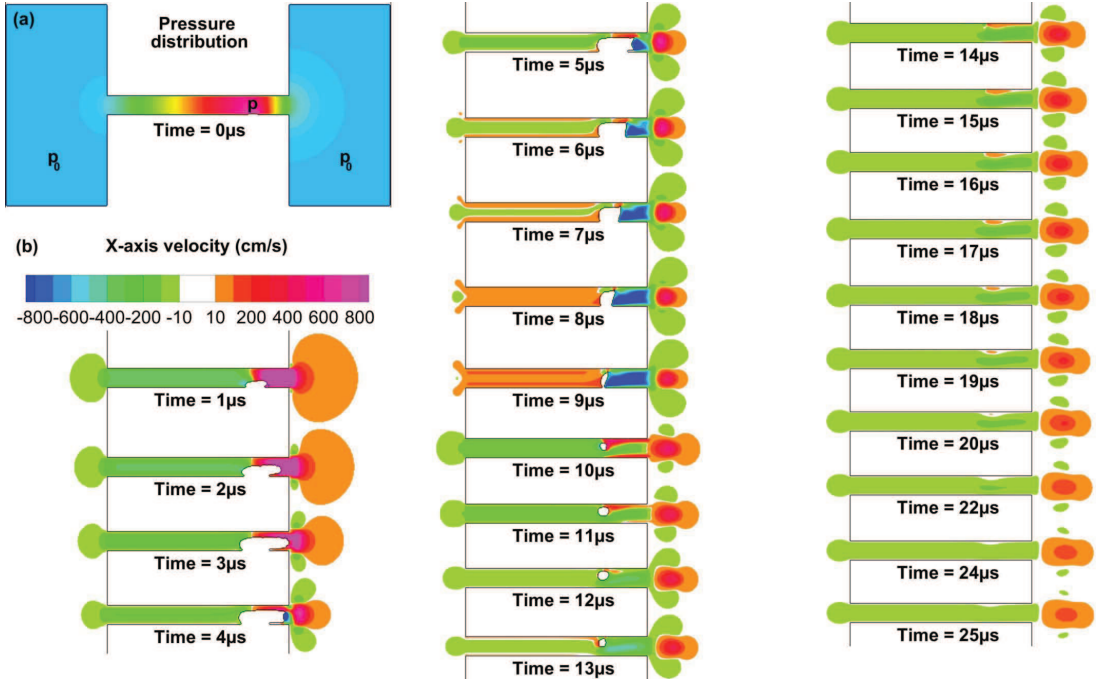}
\caption{Images of pressure distribution and axial velocity along the centerline of the channel for a single pumping event. (a) Pressure distribution at $t = 0$ $\mu$s. (b) Velocity images for 1 to 25 $\mu$s. Velocities in the range of $[-10,+10]$ cm/s are transparent.}
\label{fig:four}
\end{figure*}

In this Section, we will examine a wide range of fluid and geometric parameters important to inertial pump operation. Specifically, we look at three surface tensions (20, 50, and 70 dyn/cm) and two viscosities (1.3 and 2.6 cP). In the Section~\ref{sec:four}, a wider range of viscosity is used (1 to 15 cP) when comparing to experimental measurements. Vapor bubble strength is also varied between a normal and high strength bubbles. Geometrical parameters such as the location of the pump resistor ($R = 0.5$ to 0.95) and five channel heights are examined (10, 15, 20, 30, 40, 60 $\mu$m). The relative importance of each of these variables as well as interactions is discussed.

\subsection{\label{sec:threeone}
A single pumping event
}

A single pumping event was described qualitatively in Section~\ref{sec:two}. In this Section we show a typical example of the CFD pump simulation. First, a superheated vapor forms a thin layer above the pump resistor reaching many atmospheres of pressure near the metastable limit of the fluid. Second, the high pressure forces a rapid expansion of the vapor bubble, and subsequently the pressure in the vapor drops rapidly. Third, once the inertial expansion of the vapor bubble has been exhausted, the vapor bubble begins to contract. Finally, the vapor bubble collapses and ``rebounds'' from vapor repressurization before disappearing back into solution. Figure~\ref{fig:four} shows velocity images at the channel centerline during the vapor bubble lifetime. Note the ejection of a vortex out of the right side of the channel. This vortex slowly moves away from the end of the channel, but persists well after the vapor bubble disappears. Flow moving back into the channel when the vapor bubble collapses enters the channel around the periphery of the vortex. Flow features such as this vortex have been noted in previous works,~\cite{Wang2004,Dijkink2006,Dijkink2008} and while interesting, it is not in the scope of this paper.

\begin{figure}[t]
\includegraphics[width=0.48\textwidth]{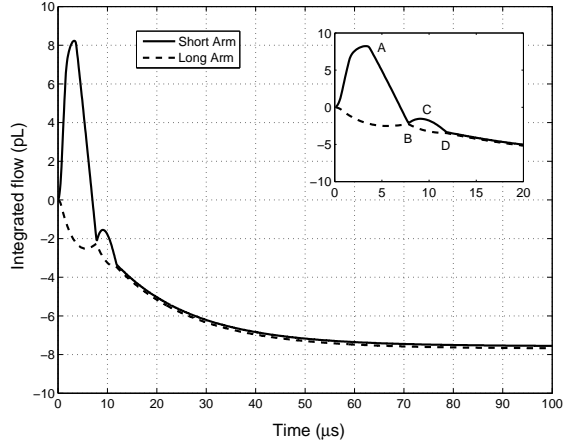}
\caption{Integrated flow volumes on both sides of a 20 $\mu$m high channel for a normal drive bubble, resistor location $R = 0.85$, 1.3 cP fluid viscosity, and 50 dyn/cm surface tension. The resistor is located in the channel as shown in Fig.~\ref{fig:three}. Inset: a magnified view of the first 20 $\mu$s. Characters indicate: A - maximum bubble extent, B - bubble collapse, C - vapor bubble rebound, D - bubble redissolving into solution.} 
\label{fig:five}
\end{figure}

In the simulation, the integrated amount of flow which passes through each end of the channel during the pumping event is tracked. Figure~\ref{fig:five} shows the integrated flow in the short and long arms of the channel during a single pumping event. Integration (measure) planes are shown in Fig.~\ref{fig:three} by solid rectangles.

During the vapor bubble expansion, fluid flows away from the vapor bubble out of the long and short arms of the channel. Since flow in the $x$ direction is positive and flow in the negative $x$ direction is negative, the integrated flow in the long arm is negative initially while the flow in the short arm is positive. After a short time (about 4.5 $\mu$s, see A in Fig.~\ref{fig:five} inset), the vapor bubble reaches its maximum extent. Subsequently, the bubble contracts and fluid reverses its flow on both sides (but not simultaneously). This causes the integrated flow in the long arm to increase temporarily as the flow is now in the positive $x$ direction, while the short arm flow becomes negative. 

The vapor bubble collapses around 8 $\mu$s (B in Fig.~\ref{fig:five} inset). At this point, the net flow is 2.1 pL meaning the collapse has shifted 2.1 pL toward the long arm, i.e. in the negative $x$ direction. After the vapor bubble rebounds, which creates a short jump in the curves at approximately 10 $\mu$s (C), the vapor bubble disappears (D). The channel continues to pump fluid in the direction of the long arm after the bubble is gone because of the residual momentum in the channel. After roughly 50 $\mu$s, the momentum is viscously dissipated and the fluid comes to rest after pumping nearly 8 pL in the long arm direction.

The flow in this last stage decays exponentially with time. According to Eq.~(\ref{eq:four}), the exponent is given by $\kappa/\rho A$. Since the fluid density and channel cross-section are known, by fitting the CFD flow to an exponential the effective viscosity $\kappa$ of the one-dimensional model can be extracted.

\subsection{\label{sec:threethree}
Effect of resistor location
}

The effectiveness of the pumping is largely determined by the location of the pump resistor within the channel. If the pump resistor is located in the center, there is no pumping effect because of symmetry. As the pump is moved toward either end, the pumping effect grows until the vapor bubble vents into the reservoir and the effect drops off. 

\begin{figure}[b]
\includegraphics[width=0.48\textwidth]{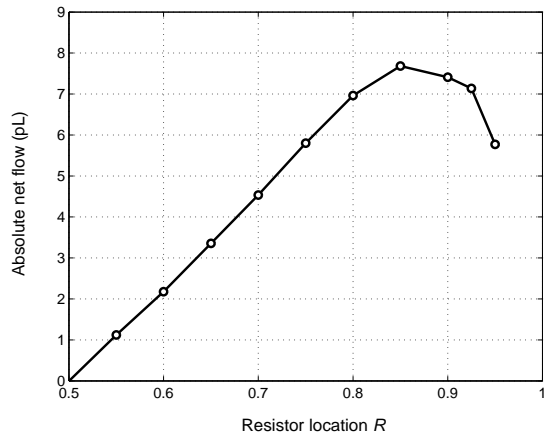}
\caption{Net flow versus resistor location in the channel for the same conditions as in Fig.~\ref{fig:five}. A resistor location of $R = 0.5$ indicates a pump in the center of the channel and $R = 1.0$ indicates a pump resistor located at the end of the channel.}
\label{fig:six}
\end{figure}
\begin{figure}[t]
\includegraphics[width=0.48\textwidth]{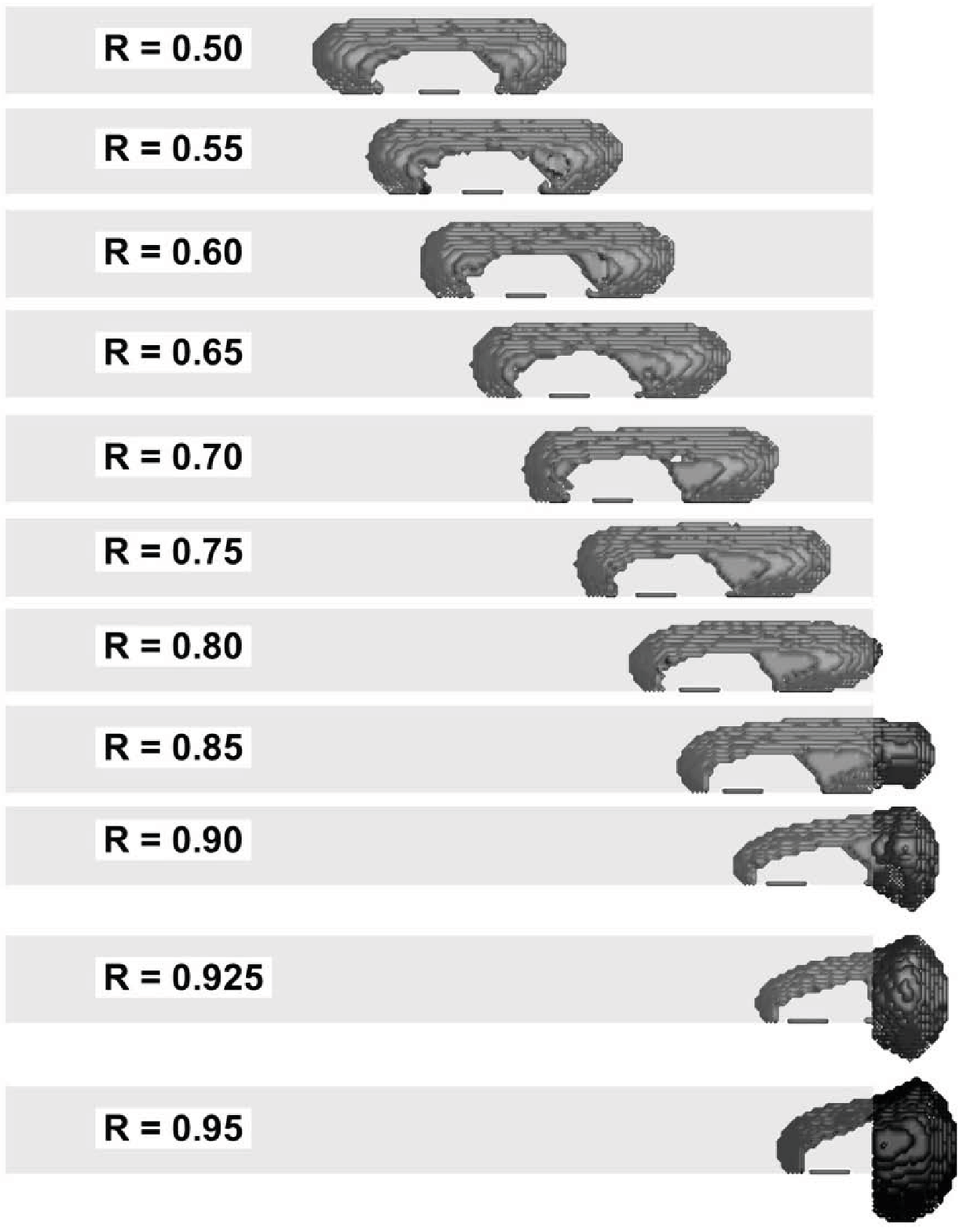}
\caption{Bubble isosurfaces at maximum extent for different resistor locations $R$.}
\label{fig:seven}
\end{figure}

Figure~\ref{fig:six} shows the effect of pump location $R$ on the absolute net flow for the same conditions as in Fig.~\ref{fig:five}. One can observe that for a resistor located in the center of the channel ($R = 0.5$) the net flow is zero as expected. When $R$ approaches 0.85 the net flow hits a maximum of 8 pL. At $R > 0.85$ the net flow decreases because the vapor bubble spills into the reservoir. Note that the relationship of the net flow with respect to resistor location is linear up to the maximal value. In the following Sections we will examine the geometric and fluid effects on the net flow curve.

\subsection{\label{sec:threefour}
Pumping in axial/non-axial bubble regimes
}

There are two general regimes of vapor bubble behavior that govern the effectiveness of pumping: {\em axial} and {\em non-axial}. The axial regime is when the bubble growth is in a direction that is aligned with the channel (i.e. the axial, or $x$, direction). In this regime, the vapor bubble is mostly contained within the channel during its lifetime. The non-axial regime is when the vapor bubble growth is out of the channel axis and the bubble spreads radially away from the channel. Note that the bubble can extend out of the channel end in the axial regime but it continues to grow only along the channel axis. If the bubble starts to grow outward, this is the onset of the non-axial regime. Figure~\ref{fig:seven} shows a series of images of {\em maximum} vapor bubble extent for different resistor locations.

One can see that $0.50 \leq R \leq 0.85$ correspond to axial operation while $R > 0.85$ to non-axial operation. Given that Figs.~\ref{fig:six} and \ref{fig:seven} present results for the same conditions, Fig.~\ref{fig:six} shows that net flow continues to increase until the onset of the non-axial regime. Net flow in the axial regime continues to increase because the asymmetry of the resistor location is increased even as the bubble extends beyond the end of the channel. This behavior stops when $R$ is moved far enough to the channel end that the bubble exits the channel while at high pressure and grows into the reservoir. {\em The transition between the axial and non-axial regimes is the resistor location for maximum net flow.}

\subsection{\label{sec:threefive}
Effect of channel cross-section
}

\begin{figure}[t]
\includegraphics[width=0.48\textwidth]{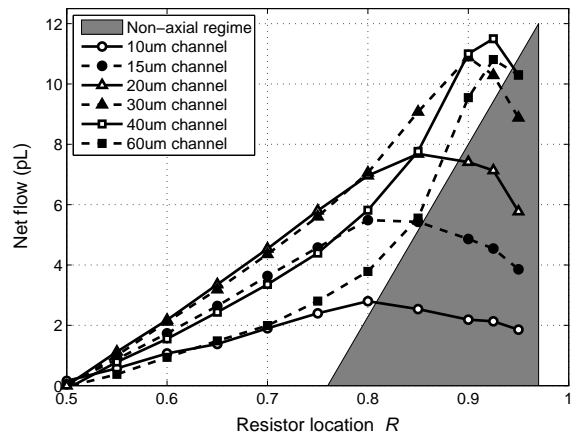}
\caption{Net flow as a function of resistor location and channel height. The shaded region indicates the non-axial regime.}
\label{fig:eight}
\end{figure}

Another important variable in the effectiveness of pumping is the channel cross-section. Different channel heights between 10 $\mu$m and 60 $\mu$m were examined, while the channel width was kept fixed at 20 $\mu$m for simplicity. Figure~\ref{fig:eight} shows the net flow as a function of channel height and resistor location, for a fluid with 1.3 cP viscosity and 50 dyn/cm surface tension using a normal vapor bubble. One can see some trends that were seen with the 20 $\mu$m tall channel. The net flow rises from zero at $R = 0.5$ up to the end of the axial regime where it hits a maximum, then the net flow decreases. 

Changing the channel's height creates a couple of differences compared to previous results. First, the locations of the axial and non-axial regions shift. Taller channels stay in the axial regime until larger $R$ than shorter channels. Second, channels with non-square cross-sections, either taller or shorter, have lower net flow in most of the axial regime.

\subsection{\label{sec:threesix}
Self-similarity in the axial regime
}

\begin{figure}[t]
\includegraphics[width=0.48\textwidth]{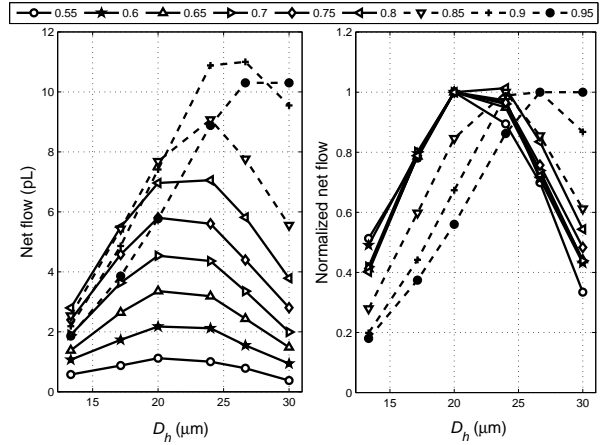}
\caption{Left panel: net flow versus hydraulic diameter for various resistor locations. Resistor location list is on top. Right panel: net flow normalized by its maximum for each resistor location. Dashed lines indicate $R$s corresponding to the non-axial regime.}
\label{fig:nine}
\end{figure}

Another useful way of visualizing the effect of channel height variations is to plot net flow versus hydraulic diameter of the channel cross-section $D_h$, cf. Eq.(\ref{eq:eightone}). This dependence has a characteristic parabolic shape as shown in Fig.~\ref{fig:nine}. The quadratic relationship breaks down as the resistor location moves to the end of the channel into the non-axial regime ($R = 0.85-0.95$ in Fig.~\ref{fig:nine}). In the non-axial regime, the taller channels with the larger $D_h$ have a larger pumping effect than the shorter channels at the same $R$. 

Because the net flow vs. $D_h$ curves in the axial regime have similar shapes, it is possible to collapse them into a single curve as shown in the right side of Fig.~\ref{fig:nine}. Here the net flow has been normalized by the maximum net flow at each $R$. As expected, resistor locations in the axial regime collapse to a single universal curve, while the ``non-axial'' locations do not. This universal quadratic function with respect to $D_h$ indicates that as the channel cross-section deviates from a square cross-section, there are higher losses in the channel reducing net flow. For shorter channels, the vapor bubble is constricted vertically and viscous losses are higher. Taller channels reduce net flow because the vapor bubble grows vertically in the channel rather than axially reducing the pumping effect.

\begin{figure}[t]
\includegraphics[width=0.48\textwidth]{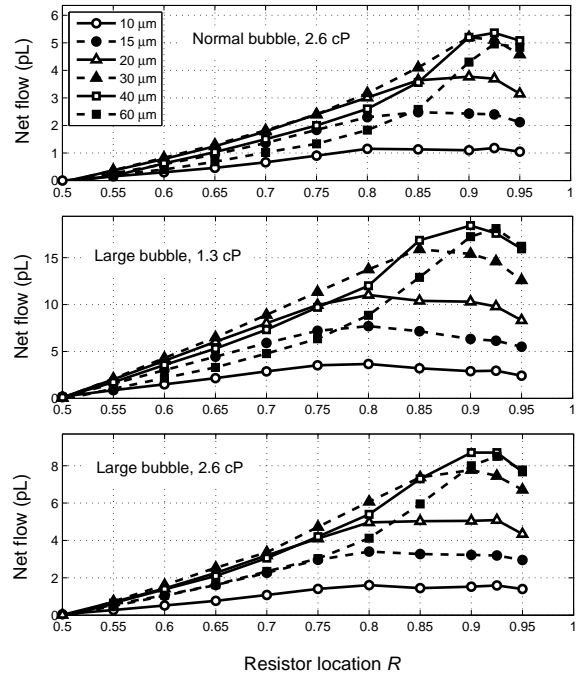}
\caption{Net flow plots for various bubble sizes and fluid viscosities. Different lines correspond to different channel heights. Top: normal vapor bubble and 2.6 cP fluid. Middle: large vapor bubble and 1.3 cP fluid. Bottom: large vapor bubble and 2.6 cP fluid. Notice different vertical scales.}
\label{fig:ten}
\end{figure}

\subsection{\label{sec:threeseven}
Changing vapor bubble strength and fluid parameters
}

The effects of changing the fluid viscosity and surface tension as well as changing the vapor bubble strength are now examined. The vapor bubble strength can be easily modified in physical experiments by changing the electrical pulse delivered to the resistor. In simulations, the vapor bubble parameters corresponding to different electrical pulses are adjusted accordingly. 

As mentioned previously, two different viscosities, 1.3 cP and 2.6 cP, and two different bubble strengths, normal and high, have been tested for a total of four cases (surface tension is 50 dyn/cm). The case for the normal bubble and 1.3 cP has already been covered and all of the resistor locations and channel cross-sections used in this case were utilized in the remaining three cases. Figure~\ref{fig:ten} shows how net flow varies as a function of resistor location for these three new cases.

\begin{figure}[t]
\includegraphics[width=0.48\textwidth]{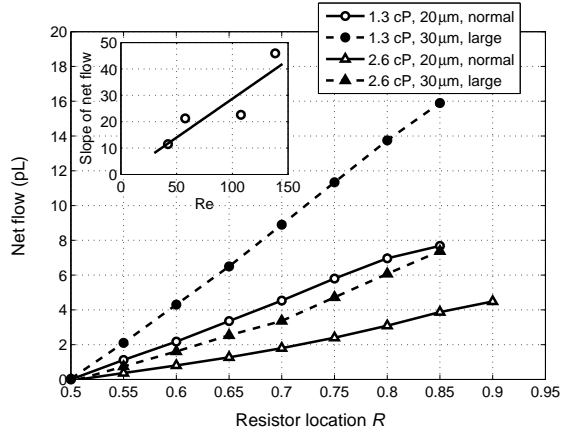}
\caption{Main panel: plots of net flow versus resistor location for four fluid and drive bubble combinations. The maximum net flow in the axial regime for each case is shown. (Respective channel heights are indicated in the legend.) `Normal' or `large' in the legend refers to the bubble strength. Inset: the slope of net flow (with respect to resistor location for the four cases in the main panel) versus Reynolds number.}
\label{fig:eleven}
\end{figure}

It is possible to analyze these new cases in a fashion analogous to the analysis performed previously for a normal vapor bubble and 1.3 cP fluid. By recasting the net flow as a function of hydraulic diameter one finds that the resistor locations in the axial regime are self-similar as before. Further, one can identify the cross-sections with the maximum net flow in the axial regime for each of the four cases, cf. Fig.~\ref{fig:eleven}. Figure~\ref{fig:eleven} shows that the net flows varies essentially linearly with $R$. Comparing the slope of each case with its corresponding Reynolds number, one can see that the slope is linearly dependent on Re (cf. the inset). This means that the inertial pump becomes more effective as the drive bubble strength increases and the viscous damping forces are reduced.

Reynolds number is defined as~\cite{White1991}:
\begin{equation}
{\rm Re} = \frac{\rho U D_h}{\mu} \: . 
\label{eq:nine} 
\end{equation}
where $D_h$ is a characteristic length (hydraulic diameter based on the channel cross section), $\rho$ is the density, $\mu$ is viscosity and $U$ is a characteristic velocity. There are several ways to define an appropriate value for $U$. In this paper, $U$ is derived from the results of CFD simulations using the formula
\begin{equation}
U = \frac{\Omega_{b}}{D^2_h \tau} \: , 
\label{eq:thirteen} 
\end{equation}
where $\Omega_{b}$ is the maximum extent of the vapor bubble, and $\tau$ is the time needed to reach maximum bubble extent. Although velocity is a function of location and time in these simulations, this velocity estimate accounts for many of the major variables including channel height, resistor location, viscosity, and surface tension. The Reynolds numbers shown in Fig.~\ref{fig:eleven} are obtained using Eqs.~(\ref{eq:nine}) and (\ref{eq:thirteen}). Because the net flow scales directly with Reynolds number, it means that pumping in the axial regime is a balance between inertial and viscous forces as described in Section~\ref{sec:two}.

\begin{figure}[t]
\includegraphics[width=0.48\textwidth]{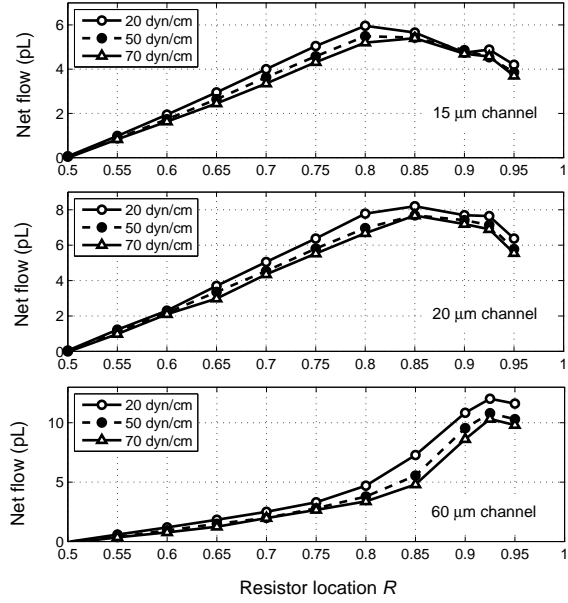}
\caption{Net flow for three different channel heights for normal drive bubbles with varying surface tension and resistor locations.}
\label{fig:sixteen}
\end{figure}

Changes in surface tension have much less effect on the net flow than viscosity variations. This is shown in Fig.~\ref{fig:sixteen}. Varying $\sigma$ from 20 dyn/cm to 70 dyn/cm has little effect on the net flow. As discussed in Section~\ref{sec:twotwo}, this is because surface tension is a relatively weak force compared to the vapor bubble pressure or the inertia of the flow during the lifetime of the bubble (i.e. the first 10 $\mu$s after boiling begins). 

One metric for assessing the relative impact of surface tension is the comparison of the average vapor bubble pressure to the capillary pressure:
\begin{equation}
f = \frac{\bar{p}_{\rm vapor}}{4\sigma/D_h} \: . 
\label{eq:fourteen} 
\end{equation}
This pressure ratio is above 100 during the initial stages of the vapor bubble lifetime, which further demonstrates the relative weakness of the surface tension. 

The relative importance of the surface forces can also be compared to inertial forces using the Weber number: 
\begin{equation}
{\rm We} = \frac{\rho D_h U^2}{\sigma} \: , 
\label{eq:fifteen} 
\end{equation}
where $U$ is the velocity from Eq.~(\ref{eq:thirteen}). For the cases shown in Fig.~\ref{fig:sixteen}, the Weber number varies from 6 to 63 indicating the larger influence of inertia relative to surface tension. From Fig.~\ref{fig:sixteen}, it is also apparent that while lower surface tensions have larger net flows, the overall effect is less than that of channel height or resistor location.

\subsection{\label{sec:threetwentyone}
Maximizing net flow
}

Based on the previous discussions, in order to maximize net flow one should maximize Reynolds number which can be achieved by adjusting fluid properties by minimizing viscosity if possible. With respect to geometry, the resistor should be placed as far to the edge of the channel as possible while staying in the axial regime. Taller channels allow resistor placement closer to the edge of the channel because they stay in the axial regime longer than shorter channels.

Previously, it was noted that for various channel cross-sections a portion of the axial regime was linear and that net flow in this regime can be scaled by Reynolds number (Fig.~\ref{fig:eleven}). Under certain conditions, there is another section of the axial regime which is superlinear, where moving the resistor location to the end of the channel increases the net flow beyond the linear rate. This occurs generally around $R \approx 0.80-0.85$ and occurs for channels 30 $\mu$m or taller because only they stay in the axial regime at these large $R$ (see Figs.~\ref{fig:ten} and \ref{fig:sixteen}). 

\begin{figure}[t]
\includegraphics[width=0.48\textwidth]{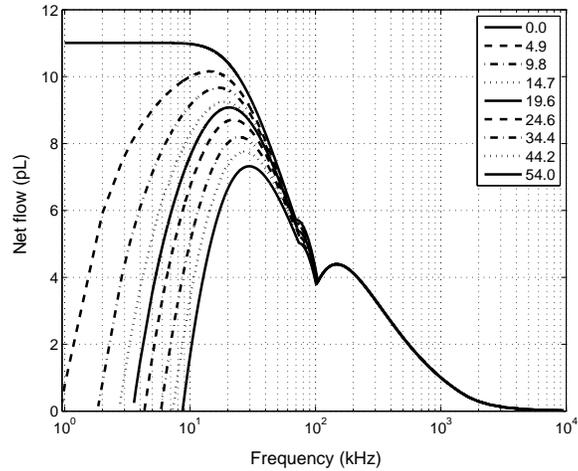}
\caption{Net flow through the channel versus frequency for a series of pressure heads. The legend numbers are pressure head values in mbar.}
\label{fig:fourteenone}
\end{figure}

The one-dimensional model of Section~\ref{sec:two} also showed this linear to superlinear transition at similar resistor locations, cf. Fig.~\ref{fig:twotwo}. Note there is no non-axial regime in a one-dimensional model (i.e. no lateral spilling of the drive bubble) so any channel height will analytically show super-linearity. The reason for this jump in net flow is that the mass in the short arm of the channel is continuously decreasing (in fact approaching zero) as the end of the channel is reached. So the vapor bubble expels fluid in the short arm into the reservoir and the bubble collapses ever more rapidly, amplifying the pumping effect as the resistor moves to the end of the channel.

\begin{figure}[b]
\includegraphics[width=0.48\textwidth]{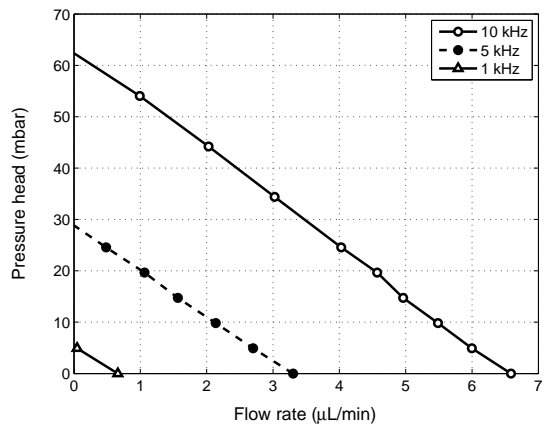}
\caption{Flow rate versus pressure head for a single pump at different frequencies.}
\label{fig:twelve}
\end{figure}
\begin{figure*}[t]
\includegraphics[width=0.96\textwidth]{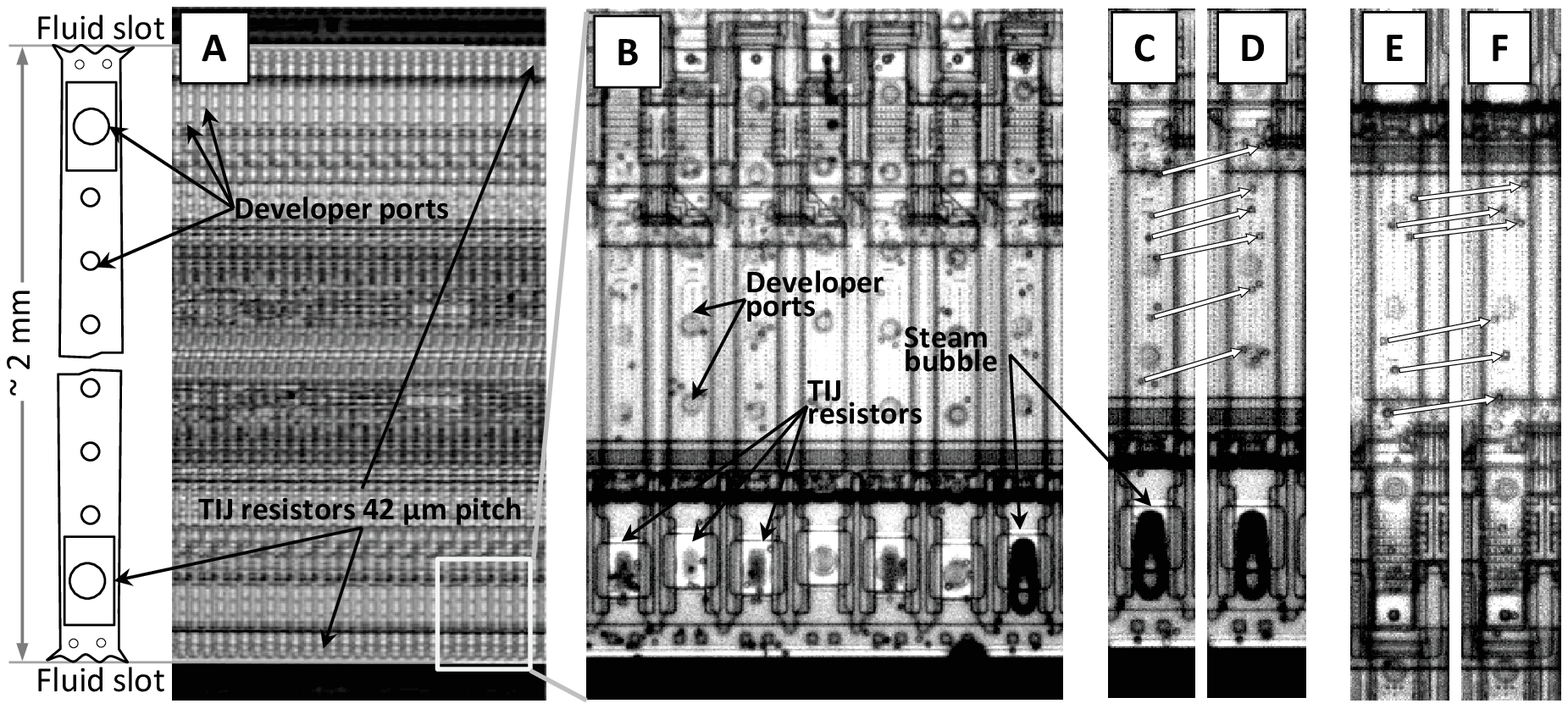}
\caption{Experimental build of 2 mm long microfluidic channels with 33 $\times$ 17 $\mu$m$^2$ TIJ micropumps. Channels' size is 25 $\times$ 17 $\mu$m$^2$ in cross-section with a 42 $\mu$m pitch. Circles are sealed developer ports required for channel processing. A: Image of 2 mm channel array. B: magnified stroboscopic image of an array of pumps demarcated by the white rectangle in A. C,D: snapshots of particles migration 400 pulses apart at 30 $^{\circ}$C. Arrows indicate migration of tracers. E,F: the same as C and D but 20 pulses apart at 60 $^{\circ}$C.}
\label{fig:fifteen}
\end{figure*}

\subsection{\label{sec:threenine}
Pump curves
}

Pump curves are used to characterize the performance of a pump to move fluid against an adverse pressure head. For the inertial pump, the pump curve is calculated by increasing the pressure in one of the reservoirs and then actuating the resistor at a prescribed frequency. Because of the transient nature of the boiling event, it is necessary to fire the resistor repeatedly in order to maintain flow. Figures~\ref{fig:fourteenone} and \ref{fig:twelve} show the characteristic performance of a single pump with a large drive bubble and 1.3 cP fluid in a 20 $\mu$m high channel with $R = 0.8$ (Re = 140, We = 5).

Notice in Fig.~\ref{fig:twelve} that the pump curves are linear. This linear relationship is expected because  Poiseuille's solution for steady, laminar flow in a pipe likely applies here (despite the transient nature of the inertial pump), and this solution shows a linear relationship of the flow rate to pressure head.~\cite{White1991} This behavior has also been found previous works for pumping in microchannels.~\cite{Wang2004,Yin2005a}

The flow rates in Fig.~\ref{fig:twelve} appear to be quite small but it should be noted that this is for a single pump. Many hundreds, even thousands of pumps may be manufactured on a single silicon die which can be combined together to increase flow considerably. In addition, the resistor size and the channel cross-section can be increased. The firing frequency of the resistor can also be increased to boost flow but up to a point as there is a resonant frequency of pumping above which it is ineffective to fire (cf. Fig.\ref{fig:fourteenone}).

\section{\label{sec:four}
Experiment
}

This Section describes experimental demonstration of inertial pumping and comparison to CFD predictions.

\subsection{\label{sec:fourone}
Device design and fabrication
}

A microfluidic system consisting of 2 mm long parallel linear channels with integrated thermal pumps was built using standard fabrication technology originally developed for Thermal Inkjet (TIJ).~\cite{Stasiak2012} The system contained 1056 parallel channels (25 $\mu$m wide and 17 $\mu$m tall), with a linear density of 600 channels per inch or 42 $\mu$m pitch (see Fig.~\ref{fig:fifteen}). Each channel had a TIJ resistor located at each end with a resistor area of 33 $\times$ 17 $\mu$m$^2$. The channels were patterned using a photoresist (SU8) on top of a silicon (Si) substrate with pre-fabricated TIJ resistors and CMOS driving electronics. The channel ends were fluidically connected to a main fluid reservoir via slots etched through the silicon substrate (the black areas at the top and bottom in Fig.~\ref{fig:fifteen}A).

\subsection{\label{sec:fourtwo}
Experimental setup and data processing
}

Manufactured parts were filled with water-glycerol mixtures of different viscosities. The goal was to characterize net flow versus viscosity in a wide range of $1-16$ cP. For high viscosity tests mixtures with up to 70\% of glycerol content were used. 

\begin{figure}[t]
\includegraphics[width=0.48\textwidth]{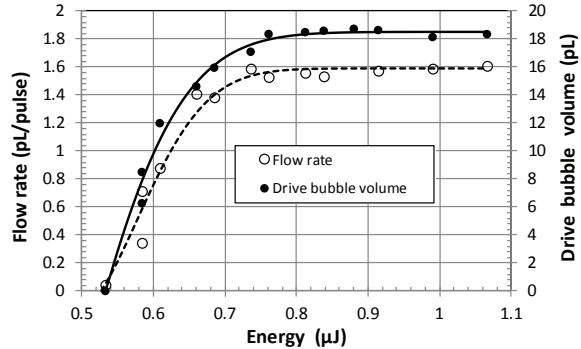}
\caption{Exemplary maximum drive bubble volume and pump flow rate vs. electrical energy stimulus.}
\label{fig:seventeen}
\end{figure}

The TIJ technology allows varying the Si die operation temperature from 28 $^{\circ}$C to 85 $^{\circ}$C with $\pm 1$ $^{\circ}$C accuracy. To estimate the firing energy, the duration of firing pulses was varied from 2.0 to 2.5 $\mu$s with the total energy per pulse in the range of $0.6-1.1$ $\mu$J. The operation energy was determined from the maximum drive bubble excursion and flow rate vs. energy curves shown in Fig.~\ref{fig:seventeen}. Both curves demonstrate a plateau (saturation) vs. firing energy of the resistor. For each test fluid and temperature, the operating energy was determined from the saturation point of the curve. In the previous Section, drive bubble strength was varied as a parameter of the CFD simulations mimicking these experimental firing energy conditions.  

The TIJ electronic driver typically enables firing frequency from 0 to 18 kHz. In experimental tests 10, 100 and 500 Hz operation frequencies were used. The flow was observed and recorded by a stroboscopic optical microscope with coaxial LED illumination of pulse duration 0.2-2.5 $\mu$s. The optical system provided magnifications from 10x to 200x, and 20x magnification was used in most tests. 

The fluid flow was visualized by adding 3.6 $\mu$m-diameter polystyrene hollow spheres as tracers. Particle image velocimetry (PIV) was used to evaluate the net flow produced by the pumps. Examples of PIV implementation are shown in Fig.~\ref{fig:fifteen} (C-D and E-F). Average particle speed was determined by measuring the migration distance per pulse. Static particles were excluded from the analysis.

\subsection{\label{sec:fourthree}
Results
}

\begin{figure}[t]
\includegraphics[width=0.48\textwidth]{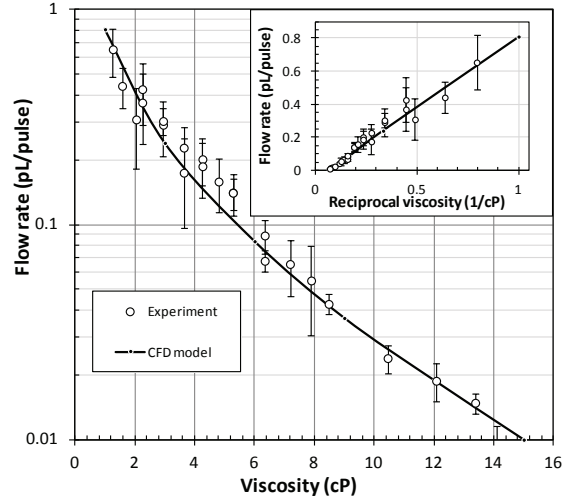}
\caption{Experimental flow rates compared with CFD results for 2 mm channel array. Inset: flow rate vs. reciprocal viscosity.}
\label{fig:fourteen}
\end{figure}

Measured flow rates as a function of fluid viscosity are plotted in Fig.~\ref{fig:fourteen}. There is excellent agreement between the experimental values and CFD simulations. If plotted versus inverse viscosity, the dependence appears to be a linear function, as shown in the inset. As expected, increasing the fluid viscosity decreases pumping performance as the fluid inertia drops relative to viscous forces.

Thermal impact of TIJ resistors was evaluated using a built-in temperature measurement capability. In one example, all 1056 pumps were fired in counter-flow mode. (In ddd and even channels fluid is pumped in opposite directions.) The measured temperature increase was less than 0.02 $^{\circ}$C per 1 Hz of operation frequency. For a starting operating temperature of 36 $^{\circ}$C up to 200 Hz and for a starting temperature of 50 $^{\circ}$C up to 1 kHz, the thermal impact was below the detectability limit ($\sim 1$ $^{\circ}$C). It should be added that, in general, the thermal impact depends not only on the operation conditions (initial operating temperature, firing energy, number of active resistors, operating frequency, etc), but also on the geometry and materials of the device.

Often there is a concern with thermal inkjet that the device may overheat or that the working fluid will be damaged or will not boil. In practice, thermal inkjet has a wide range of performance both thermally and for the variety of fluids that can be boiled. Because the heating pulse delivered to the fluid is on the order of a microsecond, the amount of fluid which experiences high temperatures is less than a micron from the resistor surface. Consequently, only a very small amount of fluid boils. Thermal inkjet has been used to jet biological material including neural cells and proteins with very little degradation.~\cite{Roth2004,Setti2005,Xu2006,Khan2010}

\section{\label{sec:five}
Summary
}

The micropump is an essential part of almost any microfluidic device. The advantages of the inertial pump presented here are the small size, mechanical robustness, and complete electronic control. As a result, the pump is versatile and reliable. Most importantly, the pumps can be made with the same materials and fabrication methods as commercial inkjet printheads. Such pumps can be made relatively cheaply at area densities of hundreds and even thousands per square centimeter. They can enable complex microfluidic networks and multifunctional devices with applications in chemistry, biology, medicine, and engineering.             

In this paper, operation of the inertial pump has been extensively studied. It has been shown that the pumping effect originates from unbalanced mechanical inertia of two columns of fluid when the actuator (a thermal resistor) is located asymmetrically inside the channel. Fluidic connection to much wider reservoirs is essential since the latter break the symmetry of one-dimensional motion by absorbing and supplying excess mechanical momentum of the fluid. {\em The pumping direction of a bubble-driven pump is always from the short arm toward the long arm of the channel,} which may appear counter-intuitive. There are two different pumping regimes: axial and non-axial. In the axial regime, the vapor bubble extends along the axis of the channel even if it is partially in the reservoirs; its transverse dimensions are equal to those of the channel. In the non-axial regime, the bubble expands transversely inside the reservoirs beyond the dimensions of the channel. The pump net flow systematically increases as the resistor is moved from the symmetrical location in the middle of the channel toward one of its ends, reaching a maximum when the axial regime gives way to the non-axial regime. For a typical 20 $\mu$m square channel with a fluid viscosity of 1 cP and surface tension of 50 dyn/cm, the flow rate is maximal at the resistor location of about 0.85. At these dimensions, a single pulse can provide about 8 pL of net flow. 

The net flow drops for non-square channels. For small channel heights the operation is hampered by increased viscous forces slowing down the fluid. In taller channels a portion of the bubble's energy is spent on expanding in the vertical direction which reduces the pump efficiency. In the axial regime, the flow curves display characteristic parabolic shapes when expressed in terms of hydraulic diameter, cf. Section~\ref{sec:threesix}. These curves can be collapsed into one universal parabola by normalizing to their respective maxima, while the flow curves of the non-axial regime do not show such universality.                 

Variation of the pump operation with fluid viscosity and bubble energy has been examined. It has been found that the flow rate scales linearly with the corresponding Reynolds number. This finding further supports the inertia imbalance as being the main factor behind the pumping mechanism. Finally, the net flow rate has been found to decrease linearly with external counter-pressure. The corresponding pumps curves are presented in Section~\ref{sec:threenine}.    

Inertial pumping has been demonstrated experimentally in a system of parallel microfluidic channels. Mixtures of water and glycerol were used to systematically vary the fluid viscosity and particle image velocimetry was used to visualize the fluid motion and measure the net flow. Excellent agreement with CFD simulations was observed. Pump operation did not result in any significant increase of the mean temperature of the device. Wide variations in fluid viscosity and in fluid types can be feasibly implemented in an inertial pump powered by a resistor with little thermal impact. In this experiment, we have used fluids with a range of viscosities (1-16 cP) and pumping performance was good.

\begin{acknowledgments}

The authors wish to thank Vlado Jakubeck and Sundar Vasudevan for fluid preparation; Tom Deskins and Tom Saksa for stroboscope assistance; Craig Olbrich and Brian Taff for TIJ die supply; Thomas Strand for SU8 technology consulting; Kenneth Abbott, Becky Angelos, Paul Benning, Tom Cooney, Christopher Davis, Mary Kent, Sue Richards, and Tim Weber for sup\-por\-t\-ing this work; Manish Giri, Vladek Kasperchik, Peter Mardilovich, Jason Oak, Vincent Remcho, and Mark van Order for useful discussions on the subject of the paper.  

\end{acknowledgments}

\providecommand{\noopsort}[1]{}\providecommand{\singleletter}[1]{#1}%

\end{document}